\documentclass[aps,twocolumn,floatfix,superscriptaddress]{revtex4}
\usepackage{epsfig}
\usepackage{subfigure}

\begin{document}

\title{Depinning and critical current characteristics of topologically defected vortex lattices}

\author{Paolo Moretti}
\affiliation{Departament de F\'{\i}sica Fonamental,
Facultat de F\'{\i}sica, Universitat de Barcelona, Av. Diagonal 647,
E-08028, Barcelona, Spain}

\author{M.-Carmen Miguel}
\affiliation{Departament de F\'{\i}sica Fonamental,
Facultat de F\'{\i}sica, Universitat de Barcelona, Av. Diagonal 647,
E-08028, Barcelona, Spain}

\begin{abstract}
We discuss the role of dislocation assemblies such as grain boundaries in the dynamic response of a driven vortex lattice. We simulate the depinning of a field-cooled vortex polycrystal and observe a general enhancement of the critical current as well as a distinct crossover in the characterisitic of this quantity as a function of pinning density. The results agree with analytical predictions for grain boundary depinning. The dynamics of grain boundaries thus proves an essential mechanism underlying the flow response of defected vortex lattices and the corresponding transport properties of the superconducting material. We emphasize the connection between the topological rearrangements of the lattice and its threshold dynamics. Our theory encompasses a variety of experimental observations in vortex matter as well as in colloidal crystals.

\end{abstract}
\maketitle

The possibility of tuning mechanical properties of vortex lattices in type II superconductors has made of vortex physics a versatile framework to study several central problems of condensed matter flow~\cite{BLA-94}. 
One of the most intriguing experimental features of vortex dynamics in disordered samples is the emergence of memory effects. Hysteresis of the $I-V$ curve is often encountered when driving defected vortex lattices above and below their depinning threshold in both low $T_c$ superconductors, such as NbSe$_2$~\cite{HEN-96,XIA-00,PAU-05} and high-$T_c$ anisotropic superconductors, such as B$_2$Sr$_2$CaCu$_2$O$_8$~\cite{SAS-00}. Numerical evidence of hysteretic behavior has also been collected, by simulating driven vortex dynamics in the presence of quenched disorder. The system was initially ``prepared" in a disordered state by letting it relax from a high-temperature liquid-like phase, according to what is know as a {\it field-cooling} protocol in experiments~\cite{MOR-05}. Afterwards currents were ramped up and down and two distinct branches could be found in the $I-V$ characteristic. The emergence of hysteresis indeed represents a shattering evidence of plastic flow in driven lattices. The voltage recorded in experiments is a direct measure of collective velocity of the vortex ensemble, while the applied currents act on vortices as an external driving force through a Lorentz-like coupling~\cite{BLA-94}. An $I-V$ measurement is thus an experimental visualization of the force vs. velocity relation in a driven over-damped medium. The elastic theory of transport in such systems, the well-know elastic depinning theory, does not account for history dependence in force-velocity relations~\cite{LED-08}. Such beyond-elastic features suggest the appearance of plastic phenomena and their involvement in memory effects. The critical current of the defected vortex array is thus a plastic threshold to vortex motion, which is found to depend on the history of the sample. Several experimental studies of the critical current $J_c$ of vortex lattices have been carried out through the past decades. They include measurements of $J_c$ as a function of both the applied magnetic field $H$~\cite{WOR-86,HEN-96} and the temperature~\cite{HEN-98}. In all systems, as a field-cooled vortex assembly is driven by a DC or pulsed signal, it behaves as if it was driven out of a metastable state through an annealing process, so that the critical current of the annealed sample proves lower than that of the initial field-cooled configuration.  Such a process is governed by a current-dependent annealing time, which diverges as the critical current is approached from above \cite{HEN-96}. In fact it was shown that in the presence of surface barriers at the edge of a sample, the field-cooled and the annealed phases can even coexist, as a result of the balance between bulk currents, which drive the annealing process, and edge currents, which produce contamination effects~\cite{PAL-00, MAR-02}. It is now widely accepted that such a non-annealed field-cooled state should correspond to a peculiar topological rearrangement of the vortex array, namely a disordered phase. This is in agreement with the experimental observation of glassy features such as power-law distributed relaxation times~\cite{HEN-96}. High-drive annealing is thus a process by which the vortex assembly recovers its topological order. The correspondence between topological disorder and higher critical currents is ubiquitous in such systems, another example being the phenomenology of the so called {\it peak effect}. In that case a critical current jump accompanies a disordering transition at high temperature or field. In field-cooling experiments, instead, one focuses on the metastable low-field and low-temperature state originated after a rapid temperature quench. 

As for the nature of the field-cooled phase, Delaunay triangulation patterns suggest that the topological order of the vortex lattice is broken by edge dislocations, which tend to arrange themselves into linear arrays such as grain boundaries (GBs). The vortex array is thus frozen in a polycrystalline state.  Vortex polycrystals have been observed, after \textit{field-cooling}, in various superconducting materials
such as NbMo~\cite{GRI-89,GRI-94}, NbSe$_2$~\cite{MAR-97,MAR-98,PAR-97,FAS-02}, BSSCO~\cite{LIU-94} and YBCO~\cite{HER-00}. 

In this Letter, we address the problem of plastic depinning of such systems. We simulate driven vortex polycrystals and demonstrate that for weak pinning forces, the threshold behavior observed in field-cooled samples agrees with our analytical predictions based on mechanisms of grain boundary depinning. We are able of  establishing a tight connection between the topology of the vortex polycrystal and its electrodynamical  response. This correspondence was first suggested in early numerical studies~\cite{FAL-96,CHA-02}. Our aim is to determine how the current response of these systems is affected by tuning crucial parameters such as the defect density $N_p$ and the magnetic field (proportional to the number of vortices $N_v$), and to emphasize how the topology of the vortex ensemble reflects these changes. Transport properties in our simulations are quantified by looking at the critical current $J_c$ of the vortex array and the collective velocity of the system in the steady state $v$ (more precisely, its component along the vortex drift direction).

Simulations are performed in a system of linear size $L=36\,\lambda$~\cite{footnote1} with periodic boundary conditions. All lengths are expressed in units of $\lambda$. In particular, we choose a value of $\xi=0.2\lambda$ for the coherence length (Ginzburg-Landau parameter $\kappa=5$, as found e.g. in certain low-$T_c$ superconducting alloys). Vortex dynamics is simulated by integrating Langevin equations of motions in the presence of quenched disorder, following the same procedure described in a previous publication~\cite{MOR-05}. Field-cooled (FC) configurations are obtained by letting a random vortex array relax in the impure environment and in the absence of external forces. The systems relaxes into a vortex polycrystal, in agreement with experimental evidence.

\begin{figure}
\centering 
\hspace{0.4cm}\subfigure[]{\epsfig{file=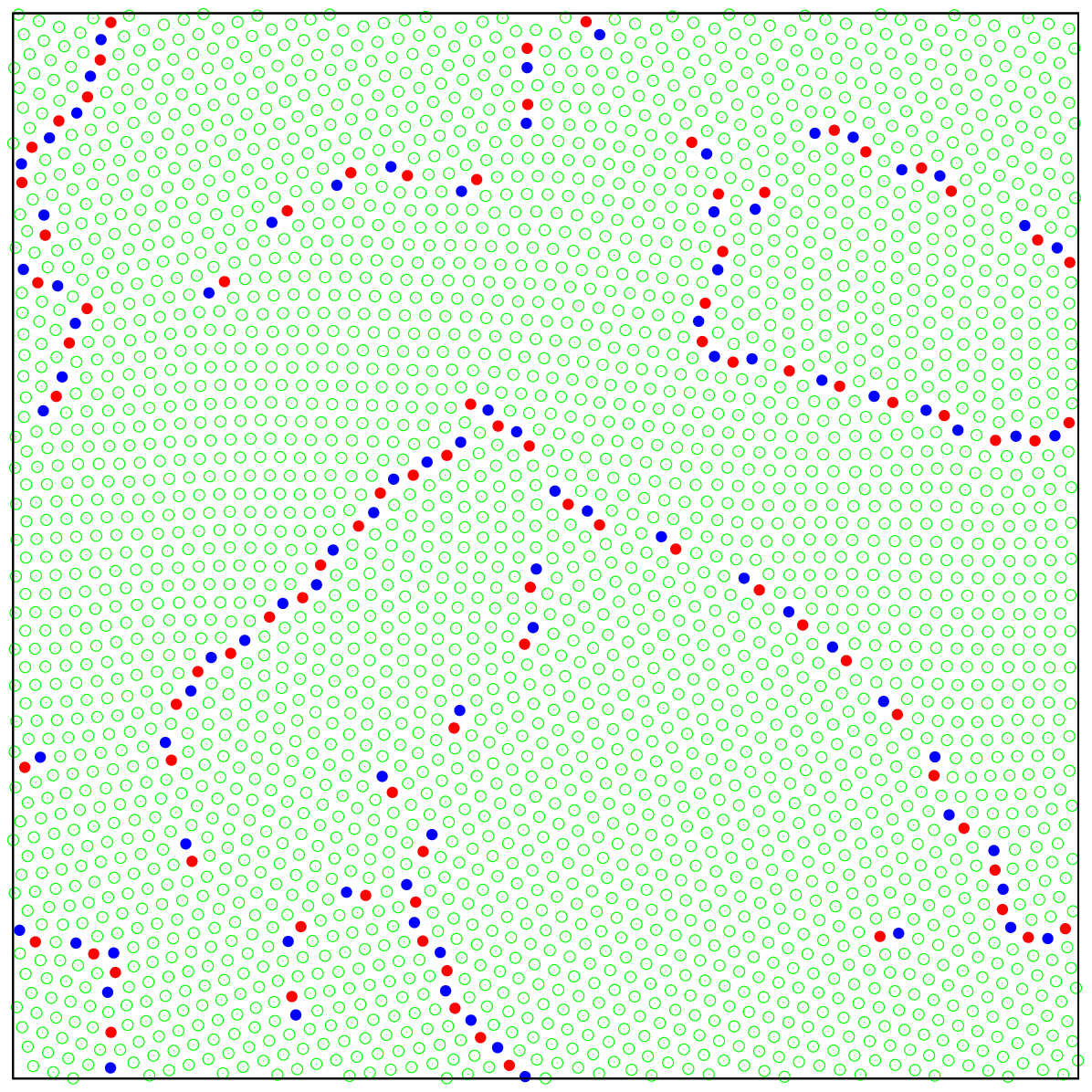,width=3.8cm,clip=}}
\hspace{0.2cm}\subfigure[]{\epsfig{file=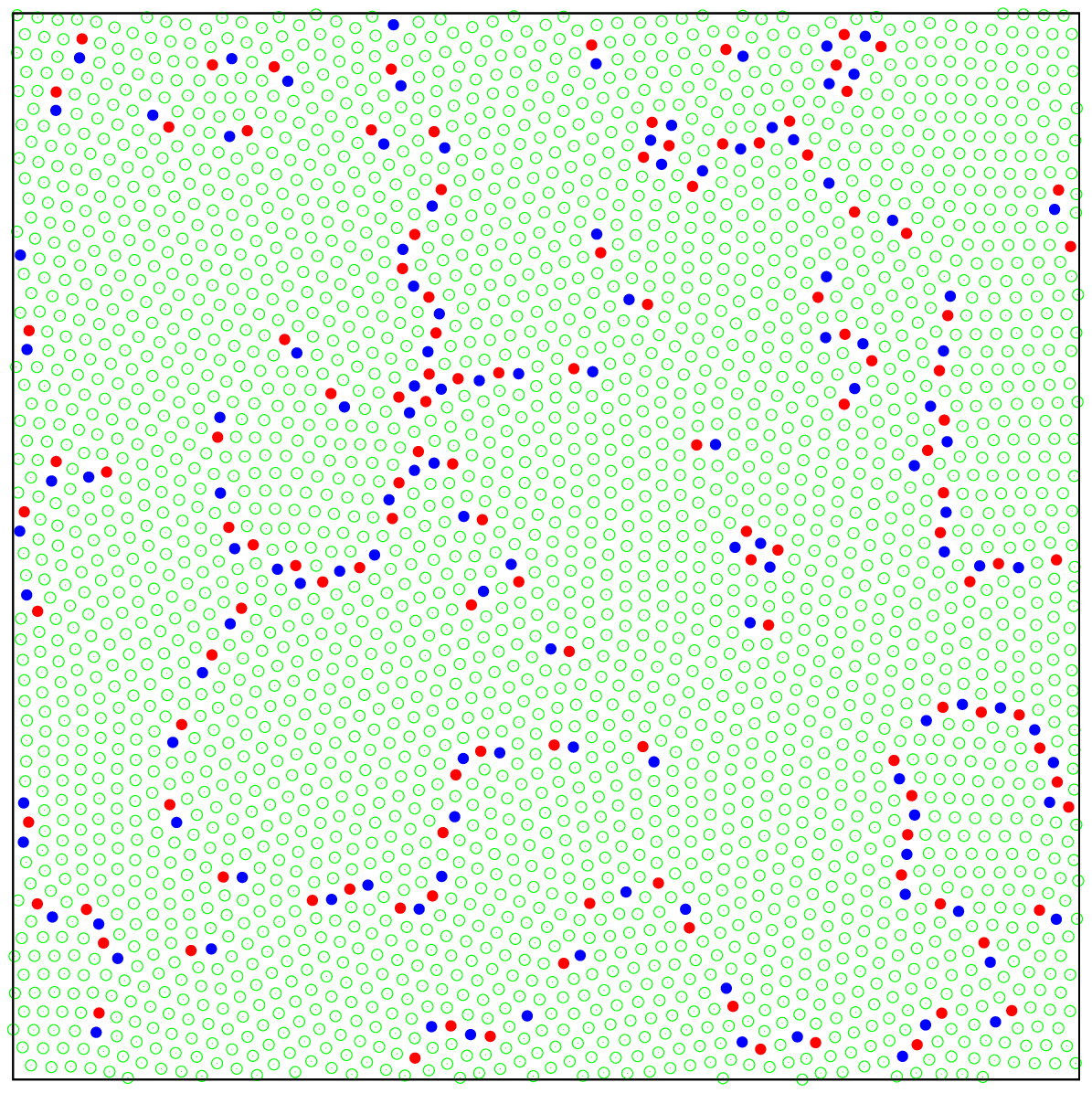,width=3.8cm,clip=}}\\ 
\subfigure[]{\epsfig{file=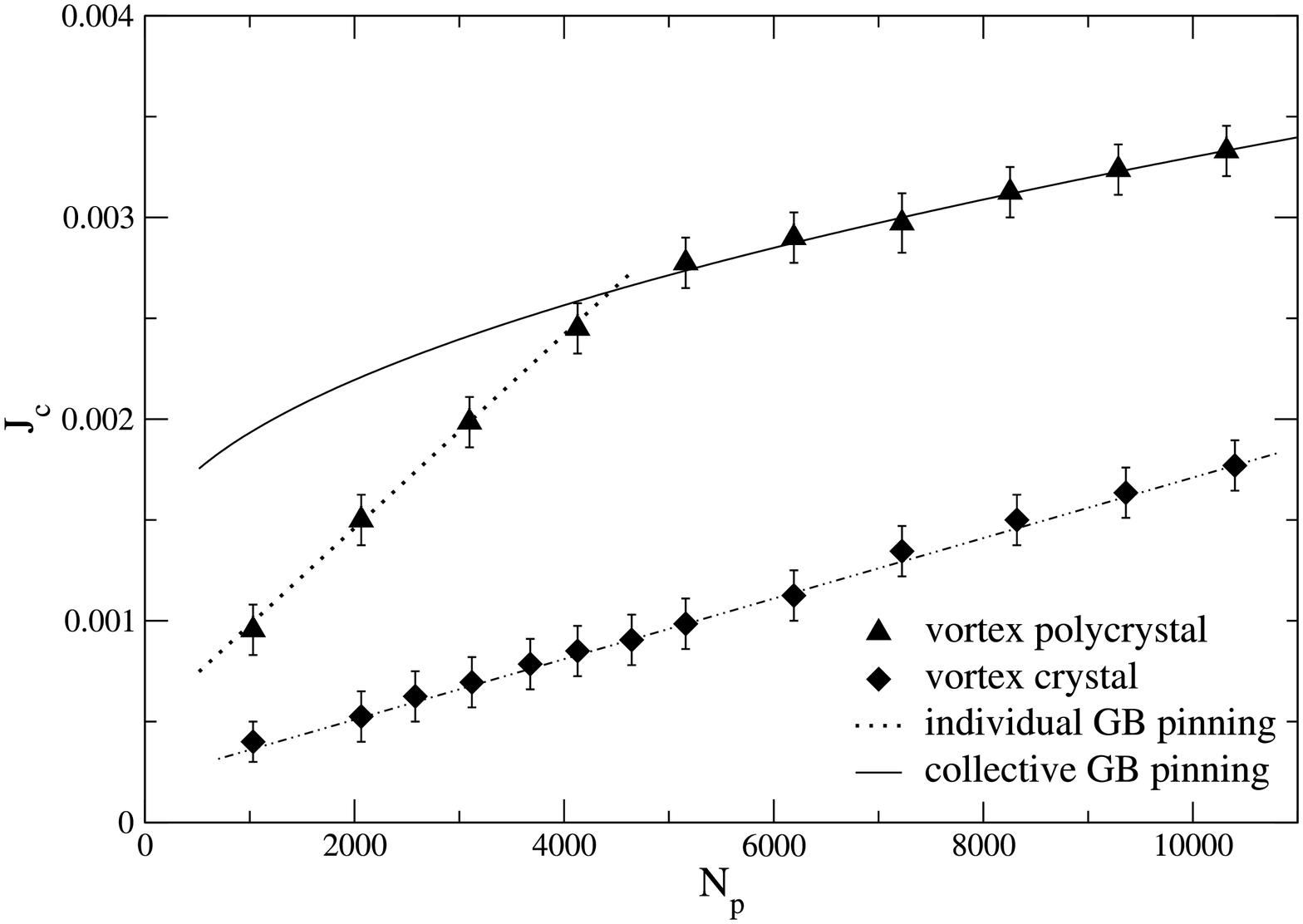,width=8.5cm,clip=}}
\caption{(a) Grain structure of the relaxed state for low pinning densities ($N_p=1032$) and (b) for high pinning densities ($N_p=8256$). Dislocations are identified as pairs of $7-$ and $5-$coordinated vortices (blue and red respectively - color online) (c) Critical current of the {\it field-cooled} vortex array as a function of the number of quenched impurities, as obtained for a vortex polycrystal (triangles) and a vortex single crystal (diamonds), for $N_v=3120$ vortices. Numerical results are compared to theoretical predictions (dotted and solid lines) for grain boundary pinning (see text). The straight line joining vortex-crystal points (dash-dotted) is drawn as a guide to the eye \cite{units}.}
\label{jcrit3_fit}
\end{figure}

We expect the number of defects $N_p$ to affect the relaxation process. As a matter of fact, defects are responsible for the polycrystalline order observed after relaxation. Vortex mutual interactions try to restore lattice order, however defects hinder this process. While varying the number of pinning centers $N_p$, we observed that all relaxed systems exhibit grain structure. However typical grain sizes are found to decrease while increasing $N_p$ only for low defect densities. Further increases of $N_p$ do not produce any appreciable drop in grain sizes. At very high $N_p$, grain structure becomes extremely complex, however it seems that grains have reached a minimum limiting size.  Figures \ref{jcrit3_fit} (a) and (b) display typical vortex array topologies in both regimes.

\begin{figure}
\centerline{\epsfig{file=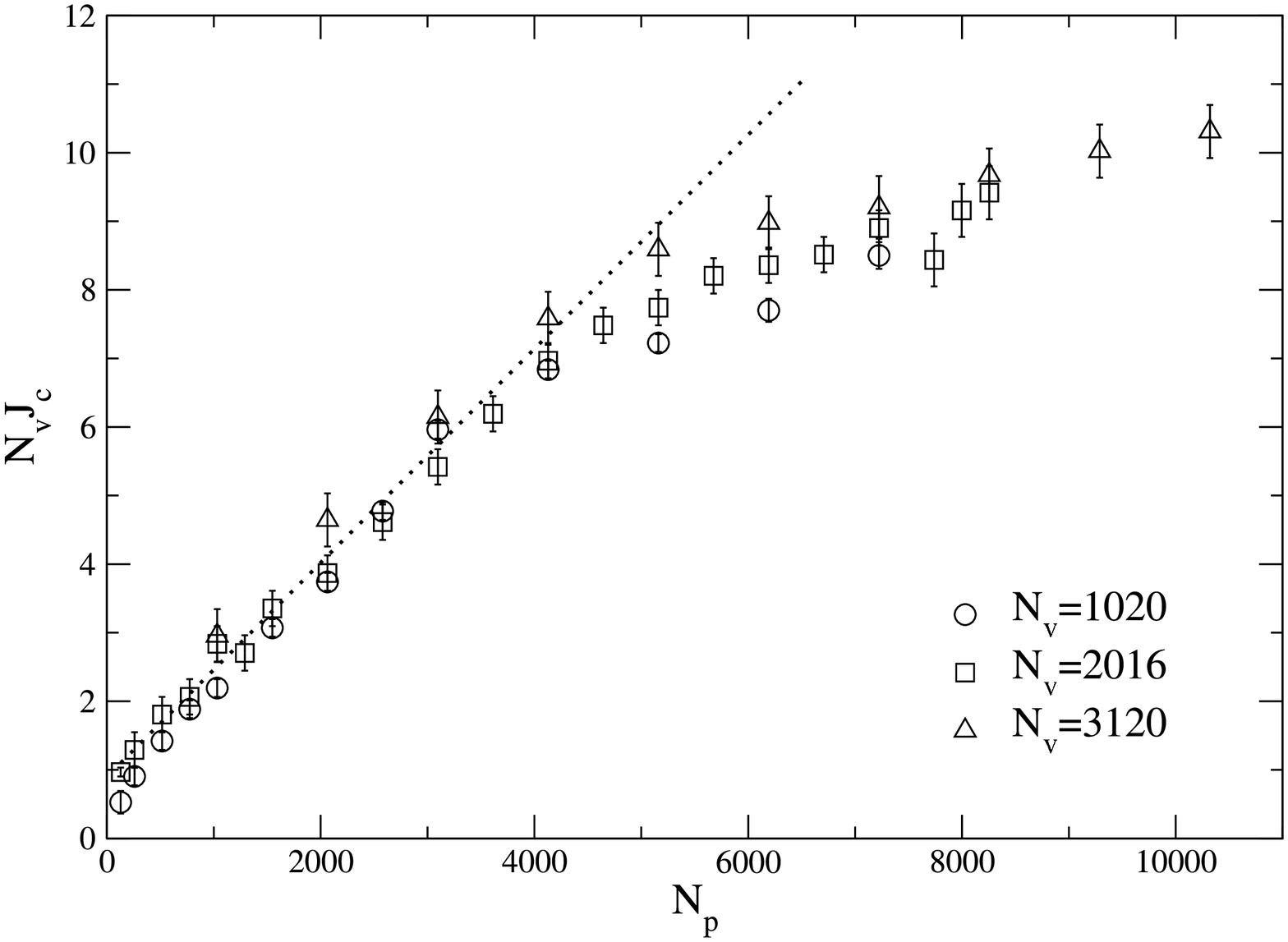,width=8.5cm,clip=}}
\caption{Rescaled critical current as a function of the number of defects for three different vortex densities. The cases of $N_v=1020$, $2016$, and $3120$ vortices are depicted. Numerical data collapse where an individual GB pinning description holds, while they deviate as soon as collective GB pinning mechanisms take over. The dotted line is drawn as a guide to the eye.}
\label{jcrit_N}
\end{figure}

We found that this behavior shows a natural correspondence with critical currents. Starting from the above relaxed configurations, we determine critical currents, as reported in Figure \ref{jcrit3_fit} (c). As expected, we see critical currents increase with increasing defect densities. At the same time, we observe that where we previously spotted a qualitative change in the grain size decrease, a crossover from a linear to a slower increase in critical currents appears.  Similar behavior is obtained for different vortex densities.  Indeed, Figure \ref{jcrit_N} shows the same critical current traits for simulations with different numbers of vortices in the system $N_v=1020$, $2016$, and $3120$.

In order to understand this correspondence, we have to focus on how grain structure might affect depinning of the vortex assembly. Compared to a perfect lattice, a polycrystal is a system with a much larger number of degrees of freedom, coming from the countless ways dislocations and grain boundaries can cooperatively move and rearrange. As a consequence, a vortex polycrystal will find it easier to adjust to disorder than a perfect lattice or even than a less defected polycrystal. A better adjustment to disorder means in general a higher depinning threshold and eventually a higher critical current. Then it is no coincidence that critical currents increase {\it rapidly} (linearly) in the region where grain sizes decrease and slow down where no further grain size decrease can be discerned. 

At this point, one might wonder what the behavior of a perfect crystal would be under the same conditions. A perfect crystal relaxes in the presence of weak disorder into the well known Bragg glass phase, which retains triangular lattice ordering over accessible length scales~\cite{KLE-01}. Such an experimental protocol is known as zero field-cooling (ZFC). We simulated ZFC lattices and measured critical currents as we did for their FC polycrystalline counterparts. Results for the case of $N_v=3120$ are also shown in Figure \ref{jcrit3_fit} (c) for comparison. Two main observations can be made: i) as discussed above and confirmed by experiments~\cite{HEN-96,XIA-00,PAU-05,WOR-86}, critical currents are higher for a vortex polycrystal; ii) no crossover behavior is observed for a vortex crystal as the number of pinning points $N_p$ is increased. The latter observation represents a crucial result of our work. The dynamic response of a vortex polycrystal proves radically different from that of a dislocation-free lattice. The presence of dislocation assemblies such as grain boundaries seems to affect the depinning of the vortex array in dramatic fashion. 

\begin{figure}
\begin{center}
\epsfig{file=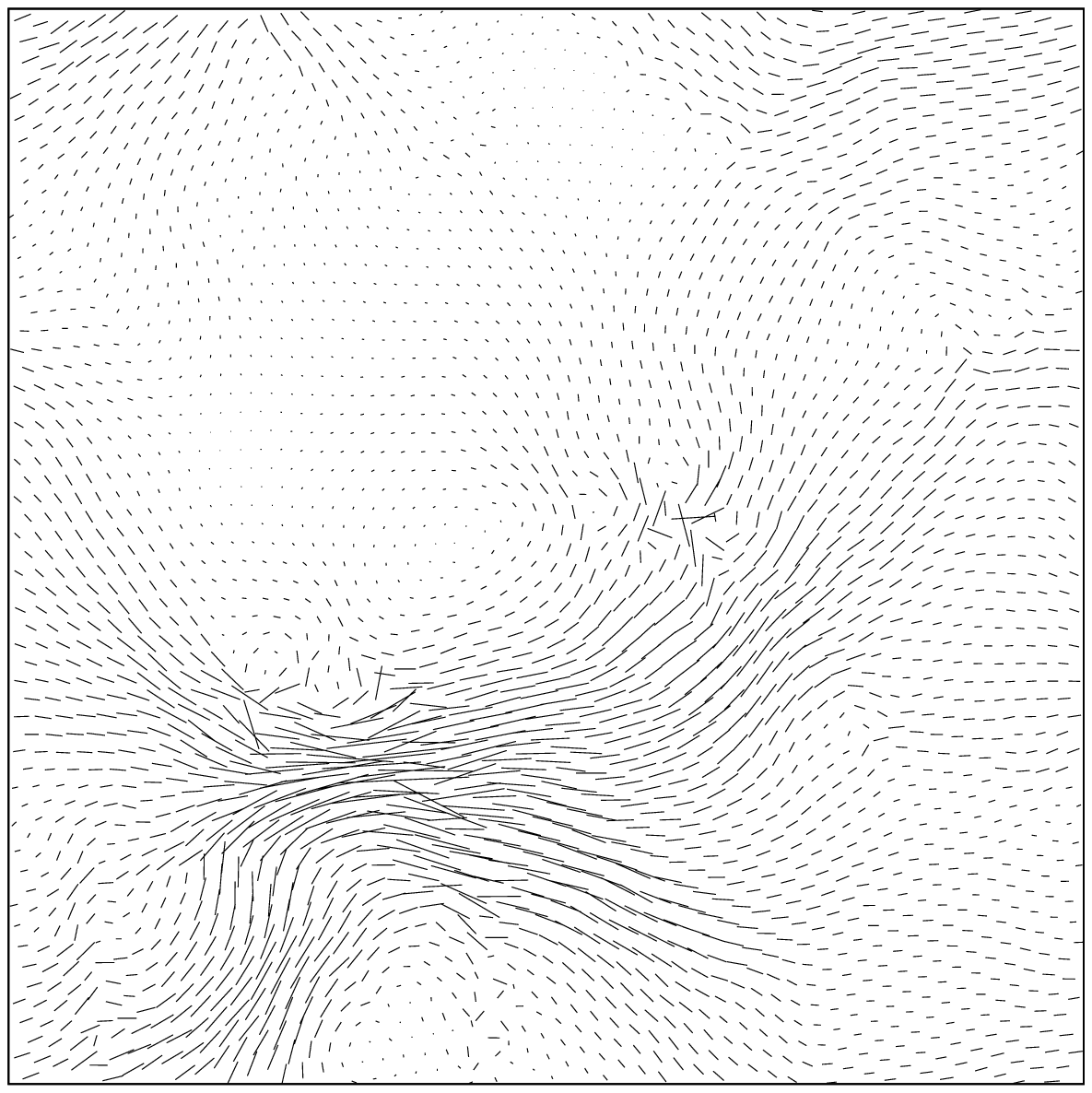,width = 3.5cm,clip=}
\hspace{0.2cm}
\epsfig{file=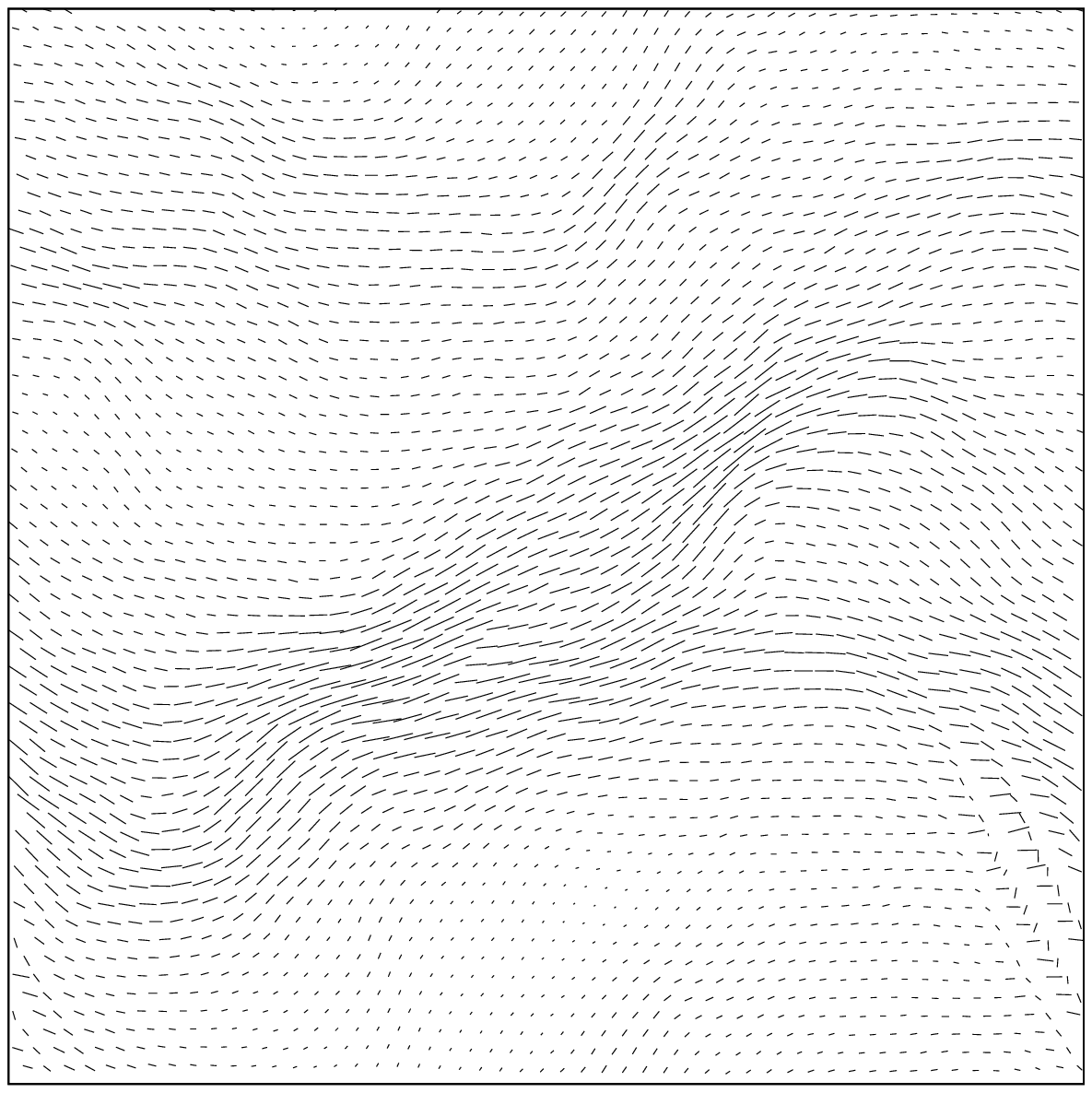,width = 3.5cm,clip=}
\end{center}
\caption{Velocity fields at incipient depinning for a FC polycrystal  (left) and a  ZFC perfect lattice (right). Plastic depinning and dislocation dynamics are signaled by the emergence of heterogeneity and convective vortex rearrangements. In a perfect vortex lattice, instead, elastic depinning is accompanied by widespread avalanches.}
\label{pvphotos}
\end{figure}

\begin{figure}
\centerline{\epsfig{file=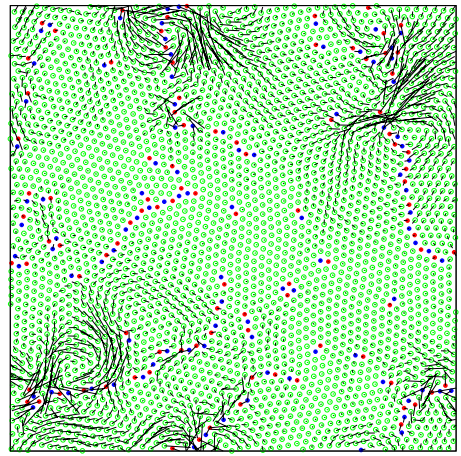,width=6cm,clip=}}
\caption{Velocity fields at incipient depinning for a FC polycrystal. The grain structure is highlighted. Vortex motion is nucleated exclusively around depinning grain boundaries. }
\label{map}
\end{figure}

We found that close to depinning the dynamics are highly heterogeneous (see Figure \ref{pvphotos}). Several swirls appear and disappear intermittently in the vortex velocity field and produce significant transversal excursions. By superimposing vortex velocity fields and the corresponding topological arrangements (Figure \ref{map}) and monitoring the time evolution of the system, we can conclude that vortex motion is activated exclusively around depinned grain boundaries and complicated velocity patterns emerge in response to GB and dislocation motion. Such regions show higher mobility and large transversal deviations. On the contrary, dislocation free regions remain dynamically frozen, unless forces well above the depinning threshold are applied. Heterogeneity and coexistence of pinned and moving regions are indeed observed experimentally in vortex matter \cite{JEN-88} as well as in colloidal polycrystals \cite{PER-08}.

Such observations corroborate the idea that grain boundaries and possibly dislocation assemblies in general are crucial in the depinning transition of vortex polycrystals. As a consequence, we propose that the threshold mechanisms of a driven vortex polycrystal should be governed by grain boundary depinning. In two recent publications of ours we proposed a theory of GB depinning~\cite{MOR-05,MOR-04b}. For high defect concentrations, the vortex lattice is collectively deformed by the superposition of strain fields produced by defects. The straining of the vortex lattice results in a collective stress field acting on vortex lattice dislocations and grain boundaries as an effective pinning field. We refer to this regime as collective GB pinning and recall that the critical stress $\sigma_c$ for GB depinning under the conditions of our simulations obeys the following relation~\cite{MOR-05}
\begin{equation}
b\sigma_c^{coll}=K\frac{D^4}{b^2}\frac{1}{R_a},
\end{equation}
where $D$ is the average dislocation spacing within the GB and  $b$ the modulus of the dislocation Burgers vector.  $R_a$ is the collective pinning length of a 2D vortex lattice and its expression is given by $R_a=c_{66}/(f_0n_p^{1/2}n_v^{1/2})$ \cite{FAL-96}, where $f_0$ is the typical pinning strength acting on the lattice, $n_p$ is the defect density, strictly proportional to $N_p$, $n_v$ is the vortex density, while $c_{66}$ is the shear modulus of the vortex lattice. $K$ is instead the shear modulus of the vortex polycrystal. Under such conditions of high pinning density, we already pointed out that the vortex lattice is highly defected and one has to assume that in principle $K \ne c_{66}$. For low defect concentrations, instead, pinning is not mediated by the lattice and grain boundary dislocations are individually pinned by impurities. The individual GB depinning stress then reads\cite{MOR-04b}
\begin{equation}
b\sigma_c^{ind}=\frac{1}{K'}\left(\frac{D}{b}\right)^2f_0^2n_p.
\end{equation}
$K'$ is the shear modulus of the vortex polycrystal for low pinning concentrations. Under such conditions, the vortex lattices is split in very large grains and one can assume that, to a  first approximation, $K'\approx c_{66}$. Replacing the correct expression for $R_a$ and focusing on the dependence on $N_p$, $N_v$ and $f_0$ we can conclude that for the critical current of  a vortex polycrystal, the following relations hold
\begin{eqnarray}
J_c\propto \left\{
\begin{array}{lc}
f_0^2\,\,(N_p/N_v) & \mbox{individual GB pinning}\\
f_0\,\,(N_p/N_v)^{1/2} K(N_v)/N_v& \mbox{collective GB pinning}
 \end{array}\label{eq:J_c}
\right.\end{eqnarray}
respectively for low and high defect concentrations. In Equations (\ref{eq:J_c}) we have exploited the proportionality of $N_v$, $c_{66}$ and the magnetic induction $B$. It is also implicit in (\ref{eq:J_c}) that the critical current $J_c$ is proportional to the 2D critical force $b\sigma_c$. We can now compare our theoretical predictions of grain boundary pinning to our numerical results for FC driven vortex polycrystals. Figure \ref{jcrit3_fit} shows how, for low defect concentrations, critical currents grow linearly with $N_p$, while for higher pinning densities $J_c$ crosses over to a square-root growth. The agreement with the theory is remarkable. At this point, we should remark that in our simulations grain boundaries develop spontaneously during the relaxation process. The fact that the global behavior matches the one analytically predicted for a GB-pinning dominated system allows us to conclude that GB pinning is the relevant mechanism that drives vortex polycrystal depinning and produces critical current anomalies, such those observed in experiments. The first equation in (\ref{eq:J_c}) also allows a prediction of the dependence of $J_c$ on the magnetic field $H\propto N_v$, for the case of low pinning density. Our theory suggests $J_c\propto 1/N_v$. Indeed, Figure \ref{jcrit_N} shows that our numerical results appear in good agreement with that prediction. Experimental curves of $J_c$ vs. $H$ for FC samples qualitatively confirm that behavior, for relatively low fields, far enough from the {\it peak-effect} region\cite{WOR-86,HEN-96}. However to our knowledge, no quantitative data are available at present. This would represent an interesting experimental test of our prediction, together with a study of the $J_c$ vs. $H$ relation in the collective pinning regime. This would unveil the physics hidden behind the $K(N_v)$ function in Equation (\ref{eq:J_c}) and ultimately allow a deeper understanding of how a complex GB network statistically affects mechanical properties of a vortex lattice.

In the light of the above observations, we can infer that plastic flow in vortex polycrystals is dominated by the gliding motion of dislocation assemblies such as grain boundaries, in analogy with nanocystalline materials. A typical fingerprint that corroborates our view is the experimental observation of broad-band noise spectra in DC-driven vortex arrays\cite{TOG-00}. Such a feature, also known as $1/f$ noise, is a natural consequence of the collective dynamics of dislocation assemblies\cite{LAU-06}, as opposed to the washboard-frequency peaks, which commonly encompass the periodicity of a dislocation-free lattice. Indeed, by means of Delaunay triangulations we observe that in our simulations, right above depinning, dislocations are created and annihilated at the same rate. At higher drives, in agreement with experiments, healing takes place in the form of a recrystallization process. After healing, by lowering the applied current the hysteresis cycle is recovered\cite{MOR-05}. Current annealing is more effective for higher vortex densities, i.e. stiffer vortex arrays, hence  the steady flow of a dense  polycrystalline vortex array along the driving direction eventually resembles the flow response of a perfect vortex crystal. On the contrary, less dense or softer arrays remain topologically defected for all the timespan of the numerical simulations.

In conclusion we performed a numerical study of plastic depinning in field-cooled vortex lattices. We observed the emergence of dislocation assemblies such as grain boundaries and investigated their role in vortex dynamics. Critical currents are found to follow the laws that we predicted analytically for grain boundary depinning processes. Grain boundary depinning thus proves the relevant mechanism accounting for the response of such systems. Our theory of GB depinning is able to explain several experimental observations like the spatial heterogeneity of the system at the threshold, the high-drive annealing of the FC state, the emergence of memory effects and the appearance of broad-band voltage noise. We established a tight connection between the topology of the vortex polycrystal and its response to externally applied currents and explored the dependance of critical currents and velocities on the density of impurities. Recent experiments on colloidal polycrystals have partially studied this dependance and provided results that agree with the predictions of our model \cite{PER-08}. We would like to propose a more systematic analysis in the same direction.      

PM acknowledges fruitful discussions with A. Kolton and M. Zaiser. This work was financially supported by the Ministerio de Educaci\'on y Ciencia (Spain), under grant FIS2007-66485-C02-02.


\begin{thebibliography}{99}
\bibitem{footnote1}
Such a length is in fact fine-tuned, in order to allow an ideal triangular vortex to fit in it.
\bibitem{units}
Current densities are measured in units of $c\Phi_0 / (8\pi^2\lambda^3)$, as in Ref. \cite{MOR-05}.
\end{thebibliography}
\end{document}